\documentclass[12pt]{iopart}

\usepackage{mathrsfs}
\usepackage{iopams}
\usepackage{cite}

\newcommand{\reff}[1]{(\ref{#1})}

\def\3nab{\tilde{\nabla}}

\def\la{\langle}
\def\ra{\rangle}
\def\c{\mbox{curl}}
\def\be{\begin{equation}}
\def\ee{\end{equation}}
\def\ba{\begin{eqnarray}}
\def\ea{\end{eqnarray}}
\newcommand{\bra}[1]{\left(#1\right)}
\newcommand{\bras}[1]{\left[#1\right]}
\newcommand{\brac}[1]{\left\{#1\right\}}

\newcommand{\sfr}[2]{{\textstyle\frac{#1}{#2}}}
\newcommand{\ffr}[2]{{\frac{#1}{#2}}}
\newcommand{\hs}{\;-\;}

\begin{document}

\title{Cosmic magnetic fields from velocity perturbations in the early
  Universe}

\author{Gerold Betschart\dag\S, Peter K.\ S.\ Dunsby\dag\ddag\ and Mattias
  Marklund\S\ }

\address{\dag\ Department of Mathematics and Applied Mathematics,
  University of Cape Town, 7701 Rondebosch, South Africa}
\address{\ddag\ South African Astronomical Observatory, Observatory
7925, Cape Town, South Africa}
\address{\S\ Department of Electromagnetics, Chalmers University of
    Technology, SE--412 96 G\"oteborg, Sweden}

\date{\today}

\begin{abstract}
We show, using a covariant and gauge\hs invariant charged
multifluid perturbation scheme, that velocity perturbations of the
matter\hs dominated dust Friedmann\hs Lema\^{i}tre\hs Robertson\hs
Walker (FLRW) model can lead to the generation of cosmic magnetic
fields. Moreover, using cosmic microwave background (CMB)
constraints, it is argued that these fields can reach strengths of
between $10^{-28}$ and $10^{-29}$ G at the time the dynamo mechanism
sets in, making them plausible seed field candidates.
\end{abstract}


\section{Introduction}

Cosmological magnetic fields have generated considerable debate
ever since they were first observed \cite{debate}. Such large
scale fields occur in galaxy clusters, high redshift
condensations, spiral and disc galaxies, and have strengths
between $10^{-7}$ to $10^{-5}\,\mathrm{G}$ \cite{Bobs}. While the
structure of the magnetic fields in spiral galaxies appear to
indicate that they were generated and sustained by a dynamo
mechanism \cite{Dynamo}, the origin of the necessary seed field is
still unknown. For the dynamo mechanism to be successful, the
present day strength of these seed fields must lie within a range
of $10^{-23}$ \cite{Kulsrud} to $10^{-30}\, \mathrm{G}$
\cite{Davis-Lilley-Törnkvist}, the lower value being for a dark
energy dominated spatially flat universe.

Primordial magnetic fields are of particular interest since they
explain both the fields seen in nearby galaxies as well as those
detected in galaxy clusters and highly redshifted condensations.
Several different schemes generating these fields have been
suggested, many of them based on breaking the conformal invariance
of electromagnetism. Since this can be achieved in more than one
way, this explains the variety of the proposed mechanisms in the
literature, such as coupling the photon to a scalar field either
during inflation or in the subsequent era of preheating~\cite{CKM}
(see \cite{GS} for a critique),  or assuming the breakdown of
Lorentz invariance either in the context of string theory and
non\hs commutative varying speed of light theories, or due to the
dynamics of large extra dimensions~\cite{BM}. Although these
proposals successfully explain the generation of primordial
fields, it is often done at the expense of simplicity because they
extend the physical parameter space.

Recently it has been suggested that if a magnetic seed field
exists below the necessary strength for the dynamo mechanism to
work, gravitational waves could be used to pump the magnetic field
by many orders of magnitude \cite{GWamp}. The mechanism uses the
primordial gravitational wave spectrum generated during inflation,
and couples this via Maxwell's equations to the magnetic seed
field. This model has the attractive feature that one need not
introduce new physics or extend the number of fields considered.
Of course, a disadvantage is that it relies on the yet unobserved
inflationary, model\hs dependent gravitational wave spectrum, and
that still requires another model to create the weak seed field.

With this in mind, we propose a mechanism based solely on the
physics of self\hs gravitating plasmas. Since velocity and density
perturbations naturally occur in the early Universe, it is
interesting to examine whether such perturbations can induce
magnetic fields which can be sustained at appreciable levels until
the onset of non\hs linear gravitational collapse (for a
discussion of density perturbations in the presence of weak
magnetic fields see,
e.g.,~\cite{Tsagas-Barrow1,Tsagas-Maartens,Hobbs}.  
The mechanism
is similar to Harrison's protogalaxy model \cite{Harrison},
and the Biermann battery effect \cite{Biermann}, in the sense of
yielding vorticity driven magnetic fields, but we note that the
battery effect in our formalism would be of second order, while
the Harrison effect relies on Thomson scattering.

Using a recently developed formalism for describing perturbations
in a general\hs relativistic multi\hs component plasma\
\cite{multifluids}, it is found that the value of the magnetic
field generated by primordial velocity perturbations is well
within the limits required  if the dynamo mechanism is to work.
The limitations of the method, as well as the results, are then
discussed.

\section{Preliminaries}

To begin with, we present some of the key equations from
\cite{multifluids} which were obtained by linearizing the exact
Einstein\hs Maxwell equations using a two\hs parameter
approximation scheme characterized by
\begin{itemize} \label{scheme}
\item $\varepsilon_g$  - gravitational: $\sigma_{ab}$, $\omega_{ab}$,
$\3nab\sigma_{ab}$,$\3nab\omega_{ab}$, etc.
\item $\varepsilon_{em}$  - electromagnetic: $E_a, B_a$.
\end{itemize}
In this way, terms which are second order in the gravitational
variables, $\varepsilon_g^2$, second order in the induced
electromagnetic fields, $\varepsilon_{em}^2$ and all cross terms
$\varepsilon_g\varepsilon_{em}$ are neglected.

\subsection{First Order Multifluid equations}

Using the above approximation scheme, the Einstein\hs Maxwell
equations are linearized around a pressure\hs free FLRW model.
Hence all spatial gradients and velocity components perpendicular
to the fundamental observer with 4-velocity $u^a$ must vanish in
the background. This implies that inhomogeneities and velocity
perturbations play a first order role. In addition, the isotropy
and homogeneity of the background demands that the electromagnetic
field and accordingly the total charge density $\rho$ vanishes in
the background.

From Maxwell's equations, we find the first order equations
\begin{eqnarray}
  \dot{E}^{\langle a\rangle}-\,\c\, B^a = -
  \epsilon_0^{-1}j^{\langle a\rangle} - \sfr{2}{3}\Theta E^a
  \ , \label{Maxwell1b} \\
  \dot{B}^{\langle a\rangle} + \,\c\, E^a =
  - \sfr{2}{3}\Theta B^a \ , \label{Maxwell2b} \\
  \tilde{\nabla}_aE^a = \epsilon_0^{-1}\rho \ ,
  \label{Maxwell3b} \\
  \tilde{\nabla}_aB^a = 0 \;, \label{Maxwell4b}
\end{eqnarray}where $\c B^a=\epsilon^{abc}\tilde{\nabla}_bB_c$ and $\Theta$ is
the cosmological expansion defined through the scale factor $S$:
$\Theta=3\dot{S}/S$

Neglecting all but electromagnetic interactions between the
fluids (e.g. collisions), then, for each fluid species (i),
conservation of energy, momentum and particle density leads to the
following equations:
\begin{eqnarray}
  \dot{\mu}_{(i)} +\mu_{(i)} \tilde{\nabla}_av_{(i)}^a=
  - \mu_{(i)} \Theta \ ,
  \label{Energy3}\\
  \mu_{(i)} \dot{v}_{(i)}^{\langle a\rangle}
  = \rho_{(i)}E^a
  - \sfr{1}{3}\mu_{(i)} \Theta v_{(i)}^a
  \ , \label{Momentum3}\\
  \dot{n}_{(i)} + n_{(i)} \tilde{\nabla}_a v_{(i)}^a =
  - \Theta n_{(i)} \ . \label{Mass2}
\end{eqnarray}
Since we are working in the energy frame, where the total heat
flux vanishes, the equation for total energy conservation is simply
given by
\begin{equation}        \label{totalenergy2}
\dot{\mu} = -\Theta\mu  \ .
\end{equation}
Because no thermal effects are included, we may express the
fluids' energy density as $\mu_i=m_in_i$ and the energy
frame choice further implies $\displaystyle{\sum_i}\mu_iv_i^a=0$.

Specializing to two components, it is convenient to introduce the
following variables
\begin{eqnarray}
  N   &=& n_1 + n_2 \ , \\
  n   &=& n_1 - n_2 \ , \\
  V^a &=& \sfr{1}{2}(v^a_1 + v^a_2) \ , \\
  v^a &=& \sfr{1}{2}(v^a_1 - v^a_2) \ .
\end{eqnarray}
The following system of first order equations then follows, using
equations\ (\ref{Momentum3}) and (\ref{Mass2}):
\begin{eqnarray}
\label{pert1}
  \dot{n}  &= -N\tilde{\nabla}_a v^a - \Theta{n} \ , \\
\label{pert2}
  \dot{N}  &= N\ffr{\Delta\mu}{\mu}\tilde{\nabla}_a v^a
    - \Theta N \ , \\
\label{pert3}
  \dot{v}^{\la a\ra} &=
    -\ffr{e}{2}\left( \ffr{1}{m_1} + \ffr{1}{m_2}\right)E^a
    - \ffr{1}{3}\Theta v^a \ ,
\end{eqnarray}
where we defined $\Delta\mu=\mu_1 - \mu_2$ and the total energy
density $\mu= \mu_1 + \mu_2$, respectively. Since we have
$\mu_i=m_i n_i$, the latter can be written as
\be \mu = \sfr{1}{2}(m_1+m_2)N + \sfr{1}{2}(m_2-m_1)n, \label{mu}
\ee
a relation which will be used very often in the following analysis. The
Raychaudhuri equation then becomes
\be
  \label{pert4}
  \dot{\Theta} = -\sfr{1}{3}\Theta^2
    - \sfr{1}{4}(m_1 + m_2)N - \sfr{1}{4}(m_1 - m_2)n \ .
\ee
It is useful to introduce the following quantities, namely
\ba
Y&=\ffr{n}{N}, \label{fraction}\\
\alpha^2 &= \ffr{4e^2}{3\varepsilon_0m_1m_2},\\
\beta^2 &=\ffr{e}{\varepsilon_0(m_1+m_2)}. \ea
The fraction $Y=n/N$ obeys a propagation equation,
\be
\ddot{Y}+\sfr{2}{3}\Theta \dot{Y} + \sfr{3}{4}\alpha^2\mu
Y=0,\label{fracprop}
\ee
which may be derived from the above definitions and equations
\reff{pert1}--\reff{pert4}.

\section{Velocity Induced Electromagnetic Fields}

For a cold plasma, the currents for each fluid species may be
written as
\be
j_{(i)}^a = q_{(i)}n_{(i)}V_{(i)}^a = q_{(i)}n_{(i)}(u^a+
v_{(i)}^a), \label{currents}
\ee
where $q_{(i)}$ is the charge and $v_{(i)}^a$ is the velocity of
the species under consideration. Since we require the plasma to be
neutral on the whole, the species are of opposite charge. Hence,
the total current $j^a$ appearing in Maxwell's equations reads to
first order
\be
j^a = j_1^a + j_2^a = -e N v^a. \label{totalcurrent}
\ee
 From Maxwell's equations \reff{Maxwell1b}
-\reff{Maxwell4b}, using \reff{totalcurrent} and \reff{pert3}, one
can then deduce second order wave equations for the induced
electromagnetic fields. They are
\ba \fl
 \ddot{E}_{\la a\ra} - \3nab^2 E_a +\sfr{5}{3}\Theta\dot{E}_{\la a\ra}
 +\bras{\sfr{2}{9}\Theta^2
 +\bra{\sfr{3}{4}\alpha^2+\sfr{1}{3}}\mu}E_a
 =2\beta^2\mu\bra{\3nab_a Y-\sfr{1}{3}\Theta
 v_a},\label{waveE} \\ \fl
  \ddot{B}_{\la a\ra} - \3nab^2 B_a +\sfr{5}{3}\Theta\dot{B}_{\la a\ra}
 +\bras{\sfr{2}{9}\Theta^2 +\sfr{1}{3}\mu}B_a
 = -2\beta^2\mu \,\c\, v_a. \label{waveB}
\ea
Observe that $B_a$ and $\c\,v_a$ are both purely solenoidal,
whereas $\3nab_a Y$ has no solenoidal  part. It is worthwhile to
note that the magnetic field is solely sourced by inhomogeneities
in the velocity in contrast to the electric field which is sourced
by inhomogeneities in the number density and velocity
perturbations. Both equations look strikingly similar, the
differences originating either from the total current or from a
gradient in the charge density (in the case of $\3nab_a Y$). The
additional $3/4\alpha^2$-term in the electric wave equation comes
from the non-stationarity of the total current and its largeness
-- $\alpha^2 \sim 10^{42}$ for an $e^+e^-$\hs plasma -- leads
directly to the high-frequency behaviour of plasma effects, as
will be shown below (see also \cite{multifluids} for a discussion
of the high-frequency plasma mode in the gravitational instability
picture).

It will be useful to introduce expansion normalized variables,
\be
  \mathscr{E}_a \equiv \ffr{E_a}{\Theta},\qquad \mathscr{B}_a \equiv
  \ffr{B_a}{\Theta},\qquad \mathscr{K}_a \equiv
  \ffr{\c\,v_a}{\Theta}.\label{exp}
\ee
Equations \reff{waveE} and \reff{waveB}, together with equations
for the driving terms, then read
\ba
\fl  \ddot{\mathscr{E}}_{\la a\ra} - \3nab^2 \mathscr{E}_a
+\bra{\Theta-\ffr{\mu}{\Theta}}
  \dot{\mathscr{E}}_{\la a\ra}  - \bras{\ffr{1}{9}\Theta^2
 - \bra{\ffr{3}{4}\alpha^2+\ffr{1}{3}}\mu}\mathscr{E}_a
 = 2\beta^2 \ffr{\mu}{\Theta}\bra{\3nab_a Y-\ffr{1}{3}\Theta v_a}
 ,\label{wavecalE} \\ \fl
  \dot{v}_{\la a \ra} + \ffr{1}{3}\Theta v_a = -\ffr{3}{8}
  \ffr{\alpha^2}{\beta^2} \Theta \mathscr{E}_a
    \label{vel}
     ,\\ \fl
  \ddot{\mathscr{B}}_{\la a\ra} - \3nab^2 \mathscr{B}_a +\bra{\Theta
  -\ffr{\mu}{\Theta}}\dot{\mathscr{B}}_{\la a\ra}
 - \bra{\ffr{1}{9}\Theta^2 - \ffr{1}{3}\mu}\mathscr{B}_a= -2\beta^2\mu
  \,\mathscr{K}_a \;,
 \label{wavecalB}\\ \fl
 \dot\mathscr{K}_{\la a \ra}  +
 \bra{\ffr{1}{3}\Theta-\ffr{1}{2}\ffr{\mu}{\Theta}}\mathscr{K}_a
 = \ffr{3}{8}\ffr{\alpha^2}{\beta^2}\bras{\dot\mathscr{B}_{\la a \ra}+
 \bra{\ffr{1}{3}\Theta-\ffr{1}{2}\ffr{\mu}{\Theta}}\mathscr{B}_a}.
 \label{Kdrive}
\ea
Equation \reff{Kdrive} follows from \reff{exp} using \reff{vel}
and Maxwell's equation \reff{Maxwell2b}. In order to find
solutions to the above equations, we extract from them the scalar
and solenoidal (vector) parts (cf. \ref{split}) and solve them separately.

\subsection{Scalar modes}

In analogy with \reff{X}, we set $V \equiv S\3nab^av_a$ and $\mathscr{%
E} \equiv S\3nab^a\mathscr{E}_a$. Equation \reff{vel} then transforms
into
\be \dot{V} + \ffr{1}{3}\Theta V = -\ffr{3}{8}
\ffr{\alpha^2}{\beta^2} \Theta \mathscr{E} = \ffr{3}{4}\alpha^2\mu S
Y, \label{vels} \ee
where the last equality is a direct consequence of Maxwell's
equation \reff{Maxwell3b}. Combining $\dot{Y}=-V/S$ with
\reff{vels} and using \reff{totalenergy2} and \reff{pert4} together
with
\be
S\3nab^a\3nab^2\mathscr{E}_a = \3nab^2\mathscr{E}
+\bra{-\sfr{2}{9}\Theta^2 +\sfr{2}{3}\mu}\mathscr{E}\;,
\ee
one can show that the scalar part of the electric wave equation
\reff{wavecalE} reduces to
\be
\ddot{\mathscr{E}} +\bra{\ffr{4}{3}\Theta-\ffr{\mu}{\Theta}}
\dot{\mathscr{E}} +\bras{\ffr{2}{9}\Theta^2
 +\bra{\ffr{3}{4}\alpha^2 -\ffr{1}{2}}\mu}\mathscr{E}
 = 0 .\label{wavecalEs}
 \ee
It is also easy to see that equation \reff{vels} additionally
gives rise to propagation equations for $V$ and $Y$:
\ba
\ddot{V}+\ffr{1}{3}\Theta\dot{V}+\bras{-\ffr{1}{9}\Theta^2
+\bra{\ffr{3}{4}\alpha^2-\ffr{1}{6}}\mu}V =0, \label{vprop} \\
\ddot{Y}+\ffr{2}{3}\Theta \dot{Y} +\ffr{3}{4}\alpha^2\mu Y=0.
\label{Yagain}
\ea
Hence, equations \reff{wavecalEs}--\reff{Yagain} all emanate from
\reff{vels}. We now specialise our considerations to a flat FLRW
background with zero cosmological 
constant, for which $\mu=1/3\,\Theta^2$ and
$\Theta=2/t$ always hold, solutions may easily be obtained:

\ba
\fl V(\tau) = \ffr{1}{\sqrt{\tau}}\brac{V_i \cos(\omega
\ln\tau)+\ffr{1}{\omega}\bra{\ffr{1}{2}V_i + V_i'} \sin(\omega \ln\tau)},
\label{solV}\\
\fl \mathscr{E}(\tau) =
-\ffr{4}{9}\ffr{\beta^2}{\alpha^2}\ffr{1}{\sqrt{\tau}}%
\brac{\bra{2V_i + 3V_i'}\cos(\omega
\ln\tau)+\ffr{(2-18\alpha^2)V_i + 3V_i'}{6\omega}\sin(\omega \ln\tau)},
\label{solE} \\
\fl Y(\tau)
=\ffr{t_i}{3S_i}\ffr{1}{\alpha^2}\ffr{1}{\tau^{1/6}}%
\brac{\bra{2V_i + 3V_i'}\cos(\omega
\ln\tau)+\ffr{(2-18\alpha^2)V_i + 3V_i'}{6\omega}\sin(\omega \ln\tau)}.
\label{solY}
\ea
Here, we introduced the dimensionless time-coordinate $ \tau
\equiv  t/t_i $, where $t_i$ denotes some arbitrary initial time.
The scale factor becomes now $S(\tau)=S_i\tau^{2/3}$. Initial
conditions of the velocity perturbation are chosen to be $V_i
\equiv V(1)$ and $V_i' \equiv V^{\prime}(1)$ (a prime stands for
$\partial_\tau$). The frequency of the solutions is proportional
to $\omega \equiv \sqrt{\alpha^2-1/36}$ and grows logarithmically
in time. The solutions show the same high\hs frequency behaviour that was
obtained in~\cite{multifluids}.

\subsection{Vector modes} \label{veccase}

According to \reff{Xv}, we set $\tilde{\mathscr{E}}_a \equiv
S\,\c\,\mathscr{E}_a$ etc., and obtain from equations
\reff{wavecalE}--\reff{Kdrive}
\ba \fl
\ddot{\tilde{\mathscr{E}}}_{\la a\ra} - \3nab^2\tilde \mathscr{E}_a
   +\bra{\Theta-\ffr{\mu}{\Theta}} \dot{\tilde{\mathscr{E}}}_{\la a\ra}
 +\bras{-\ffr{1}{9}\Theta^2
 +\bra{\ffr{3}{4}\alpha^2+\ffr{1}{3}}\mu}\tilde\mathscr{E}_a =
 -\ffr{2}{3}\beta^2\mu \,\tilde{v}_a
\label{wavecalEv} \\ \fl   \dot{\tilde{v}}_{\la a \ra} +
\ffr{1}{3}\Theta \tilde{v}_a= -\ffr{3}{8} \ffr{\alpha^2}{\beta^2}
\Theta \tilde{\mathscr{E}}_a
    \label{velv}     ,\\ \fl
   \ddot{\tilde{\mathscr{B}}}_{\la a\ra} - \3nab^2 \tilde\mathscr{B}_a
  +\bra{\Theta
  -\ffr{\mu}{\Theta}} \dot{\tilde{\mathscr{B}}}_{\la a\ra}
 +\bra{-\ffr{1}{9}\Theta^2 +\ffr{1}{3}\mu}\tilde\mathscr{B}_a =
  -2\beta^2\mu \,\tilde\mathscr{K}_a
\label{wavecalBv},\\ \fl
 \dot{\tilde{\mathscr{K}}}_{\la a \ra}  +
 \bra{\ffr{1}{3}\Theta-\ffr{1}{2}\ffr{\mu}{\Theta}}\tilde\mathscr{K}_a
 =
 \ffr{3}{8}\ffr{\alpha^2}{\beta^2}\bras{\dot{\tilde{\mathscr{B}}}_{\la
 a \ra}+
 \bra{\ffr{1}{3}\Theta-\ffr{1}{2}\ffr{\mu}{\Theta}}\tilde\mathscr{B}_a}.
 \label{Kdrivev}
\ea
Specialising to a flat FLRW background and performing a harmonic
decomposition (see \ref{split}), these equations become
\ba
 \mathscr{E}'' +
 \ffr{4}{3\tau}\mathscr{E}'+ \bras{\ffr{L^2}{\tau^{4/3}} +
 \ffr{\alpha^2}{\tau^2}}\mathscr{E}
 = - \ffr{8\beta^2}{9\tau^2}\,v\;,
 \label{Evflat}\\
 v' + \ffr{2}{3\tau}v =
 - \ffr{3\alpha^2}{4\beta^2\tau}\,\mathscr{E}\;,
\label{vvflat} \\
\mathscr{B}'' + \ffr{4}{3\tau}\,\mathscr{B}' +
  \ffr{L^2}{\tau^{4/3}}\mathscr{B} =
  - \ffr{4\beta^2}{3\tau^2}\mathscr{K}\;, \label{Bvflat}\\
\mathscr{K}' + \ffr{1}{3\tau}\,\mathscr{K} =
 \ffr{3\alpha^2}{8\beta^2}\bras{\mathscr{B}'+
 \ffr{1}{3\tau}\mathscr{B}}, \label{Kvflat}
\ea
where we have dropped the index $V$ [denoting that the variables
in equations\ (\ref{Evflat})--(\ref{Kvflat}) are vector harmonic
coefficients], and we have defined $\tau \equiv  t/t_i$ and used
\be
\bra{\ffr{kt_i}{S}}^2 =
\bra{\ffr{4\pi}{3}}^2\bra{\ffr{\lambda_H}{\lambda}}^2_i\ffr{1}{\tau^{4/3}}
= L^2\ffr{1}{\tau^{4/3}} \label{scale}
\ee
for the contribution of the Laplacian in \reff{wavecalEv} and
\reff{wavecalBv}, respectively. In \reff{scale}, $\lambda_H = 1/H$
is the Hubble length, $\lambda$ is the physical wavelength
associated with the comoving wavenumber $k=2\pi S/\lambda$ and the
index i stands for evaluation at initial time $t_i$. If the
wavelength of the mode is much greater than the horizon, eg.
$\lambda_i \gg \lambda_{H_i}$, we may neglect the terms containing
$L^2$. However, we can neglect that term in \reff{Evflat} and the
system \reff{Bvflat}\hs \reff{Kvflat} throughout, because the
$\alpha^2$\hs term dominates as long as $\tau L^3 \ll \alpha^3$
holds and this criterion is only violated for very late times or
ultra\hs short wavelengths. It follows that the above equations
can then be solved analytically and the general (real) solutions
are found to be
\ba \fl
  v(\tau) = C_1 +
  \ffr{1}{\sqrt{\tau}}\brac{C_2 \cos\bra{\omega\ln\tau} +
  C_3 \sin(\omega\ln\tau)}, \label{vvec} \\ \fl
  \mathscr{E}(\tau) = -\ffr{8}{9}\ffr{\beta^2}{\alpha^2}C_1
  -\ffr{2}{9}\ffr{\beta^2}{\alpha^2}\ffr{1}{\sqrt{\tau}}%
 \brac{(C_2+6\omega C_3)\cos(\omega\ln\tau) +
  (C_3-6\omega C_2)\sin(\omega\ln\tau)}, \label{Evec} \\ \fl
  \mathscr{K}(\tau) =
  \ffr{1}{\tau^{1/6}}\brac{D_1\cos(\tilde{\omega}\ln\tau) +
    D_2\sin(\tilde{\omega}\ln\tau)},
  \label{Kvec}\\ \fl
  \mathscr{B}(\tau) = D_3\ffr{1}{\tau^{1/3}}
  +\ffr{8}{3}\ffr{\beta^2}{\alpha^2}\ffr{1}{\tau^{1/6}}%
  \brac{D_1\cos(\tilde{\omega}\ln\tau)+
  D_2\sin(\tilde{\omega}\ln\tau)},
  \label{Bvec}
\ea
where $\omega =\sqrt{\alpha^2-1/36}$ and
$\tilde{\omega}=\sqrt{\alpha^2/2-1/36}$. Since we think of
the electromagnetic fields as being generated by the velocity
perturbations, we choose initial conditions for $v$ and $\mathscr{%
K}$, respectively: namely, $v_i \equiv v(1)$, $v_i' \equiv v'(1)$
and $\mathscr{K}_i \equiv \mathscr{K}(1)$, $\mathscr{K}_i' \equiv
\mathscr{K}'(1)$. Notice that the system of equations
\reff{Evflat} and \reff{vvflat} is equivalent to a third order ODE
for $v$ (or $\mathscr{E}$), for any value of $L$, while the system
of equations \reff{Bvflat} and \reff{Kvflat} is equivalent to a
second order ODE for $\mathscr{K}$, for $L=0$ only. However, the
vector modes of $\mathscr{K}_a$ are linearly related to those of
$v_a$ [according to (\ref{exp})], thus we have
$S\Theta\mathscr{K}\sim v$. Therefore consistency requires the
vanishing of the integration constant $C_1$. For this set of
initial conditions, the integration constants $C_i$ and $D_i$
become
\be
  C_2 = v_i, \quad C_3 = \displaystyle\ffr{v_i +
    2v_i'}{2\omega}, \quad
  D_1 = \mathscr{K}_i, \quad D_2 = \displaystyle\ffr{\mathscr{K}_i +
    6\mathscr{K}_i'}{6\tilde{\omega}},
\ee
while there is no restriction for $D_3$, since we put $L \simeq
0$, but with the assumption that the initial magnetic field
vanishes, we get a relation between $D_3$ and $\mathscr{K}_i$. The
solutions for the velocity perturbation and the (expansion
normalized) electric field agree then with those found in the
scalar case and show the same behaviour in time, as expected. The
induced expansion normalized magnetic field attains two parts, a
standard decaying part and a weakly decaying oscillatory part due
to the plasma.

\section{Applications}
\subsection{Generated fields}

Observe that the magnetic field
\reff{Bvec} is rather slowly decaying and therefore still could play a role in
some astrophysical processes under favourable conditions. If we assume that the
velocity induced magnetic field vanishes initially, then we may approximate its
magnitude by the following expression:
\be
|\mathscr{B}(\tau)| \lesssim \ffr83 \ffr{\beta^2}{\alpha^2}\ffr{1}
    {\tau^{1/6}}
\bra{1-\ffr{1}{\tau^{1/6}}}\mathscr{K}_i,
\ee
where $\mathscr{K}_i$ is the magnitude of the initial velocity curl
perturbation. Restoring SI
units, we find for the physical magnetic field
\be \fl |B(t)| \lesssim  \bra{\ffr{m_1}{m_e}}
\bra{\ffr{m_1}{m_e}} \bra{\ffr{m_e}{m_1+m_2}} \mathscr{K}_i
 \bra{\ffr{t_i}{t}}^{1/6}
\bras{1-\bra{\ffr{t_i}{t}}^{1/6}} \ffr{1}{t} \times 2\times
10^{-7}\,\mathrm{G},
 \ee
or alternatively, using $1+z = (1+z_i)/\tau^{2/3}$,
\be
 |B(z)| \lesssim
\mathscr{K}_i h
\bra{\ffr{1+z}{1+z_i}}^{1/4} \bras{1-\bra{\ffr{1+z}{1+z_i}}^{1/4}}
(1+z)^{3/2} \times 10^{-24}\,\mathrm{G}, \label{magfieldstrength}
\ee
where we also have employed $\Theta = 3H = 3H_0(1+z)^{3/2}$ with
$H_0=h(9.8\, \mathrm{Gyrs})^{-1}$, and neglected the mass
factor.\footnote{The mass parameter in\
  \eref{magfieldstrength} and \reff{eind} is always 
of the order of one for an electron-positron-- or an
electron-ion--plasma.} 
Thus, velocity curl perturbations of magnitude $\mathscr{K}_i \sim
10^{-5}$  in an $e^+e^-$\hs plasma starting after decoupling, when
Thomson scattering becomes negligible ($z \sim 1000$), would
induce a magnetic field with a strength between $10^{-26}$ and
$10^{-27}\,\mathrm{G}$ at a redshift of $z \sim 100$ (see figure 1
above), a redshift well within the limits before nonlinear effects
become important.  
Redshifting to $z \simeq 10$, at the onset of the dynamo
mechanism \cite{debate} reduces the field strength to 
between $10^{-28}$ and $10^{-29}\,\mathrm{G}$.\footnote{The
constraint concerning the initial velocity perturbation
$\mathscr{K}_i$ and $v_i$, respectively, stems from the standard
CMB results \cite{cmbvel}.} 

\begin{figure}
\begin{center}
\input{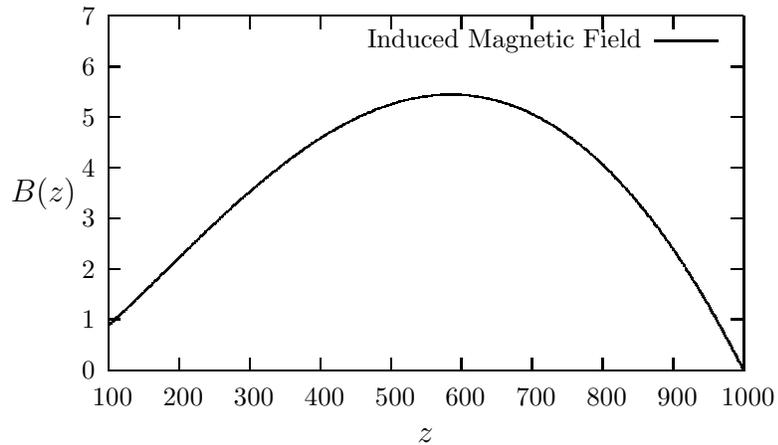}
\end{center}
\caption{Generated magnetic field in units of
  $10^{-27}\,\mathrm{G}$, for the initial and final redshift
  $1000$ and $100$, respectively.}
\end{figure}

The above argument is also applicable  for the decaying
electric field (for both scalar and vector modes). If we require the
velocity  induced electric field to vanish
initially, then its envelope is given by
\be
|\mathscr{E}_{\mathrm{env}}(\tau)| =  \ffr{4}{3}\ffr{\beta^2}{\alpha}v_i
\ffr{1}{\sqrt{\tau}}\ ,
\ee
where $v_i$ is the magnitude of the initial scalar velocity
perturbation. We resort once again to SI units and find for the
physical field
\be \label{eind}
\fl |E_{\mathrm{env}}(t)| = \bra{\ffr{m_1}{m_e}}
\bra{\ffr{m_2}{m_e}} \bra{\ffr{m_e}{m_1+m_2}} v_i
\bra{\ffr{t_i}{t}}^{1/2}
 \ffr{1}{t} \times 2\times 10^{-3}\,\mathrm{V m^{-1}},
\ee
or alternatively,
\be
 |E_{\mathrm{env}}(z)| =
v_ih
  \ffr{(1+z)^{9/4}}{(1+z_i)^{3/4}}
 \times 2\times 10^{-23}\, \,\mathrm{V m^{-1}}.
\ee
Hence, using the same data as above, i.e., velocity perturbations
of magnitude $v_i \sim 10^{-5}$ in an $e^+e^-$--plasma starting
after decoupling ($z \sim 1000$), would lead to an induced
electric field of strength $\sim 10^{-26}\,\mathrm{Vm^{-1}}$ at
$z=100$.

Comparing the energy density of the induced electric and magnetic
fields, we find that $E_{\mathrm{env}}^2/c^2B^2 \sim 10^{-15}$,
thus showing the well known fact that the electric field contribution,
due to Debye shielding, is negligible in a cosmological context.

\section{Summary and discussion}

In this paper, we have considered the induced electromagnetic
fields due to velocity perturbations in a charged multifluid,
using a gauge\hs invariant, covariant approach. We have, in a
self\hs consistent manner, investigated the behaviour of the
interacting fluids and electromagnetic fields in the case of a
flat FLRW background. In the matter dominated era, due to the
non\hs vanishing vorticity of the velocity perturbations in the
two\hs component fluid, we find that there is a net magnetic field
with a magnitude between $10^{-30}$ and $10^{-31}\,\mathrm{G}$
today. Since velocity perturbations are naturally occurring in the
early Universe, this field thus represents a suitable candidate
for a seed field, which could be amplified by the galactic dynamo.
Moreover, the model is self\hs consistent, and does not invoke any
other physics than general relativity and classical
electrodynamics

The notion of cosmic magnetic fields is nowadays generally
accepted, and observations seems to indicate their presence on all
scales of the Universe. On the other hand, the existing standard
models for the generation of these fields \cite{debate} all
require a seed field, and it is still somewhat of a mystery as to
where this primordial field stems from. There have through the
years been a number of suggestions for the origin of the seeds,
many of them making use of yet to be confirmed physics. At the
same time, CMB data is getting more and more detailed \cite{CMB},
and we now have a very good handle on the size of the different
types of perturbation that occur in the early Universe. Since it
is highly plausible that the plasma state is a good approximation
of the cosmological fluid at certain stages of the evolution of
the Universe, the model presented here lends itself naturally to
the analysis of a possible source of the much sought-after
magnetic seed field. In many of the classical mechanisms for
generating the galactic and extra-galactic magnetic fields, the
vorticity of the fluid plays a crucial role in the generation of
the magnetic fields \cite{Harrison, Biermann}. This, in
conjunction with the aforementioned CMB data, gives our model the
attractive feature of being not only internally consistent, but
also consistent with present day cosmological data.
On the other hand, scattering and thermal effects have been neglected,
and would, if included, surely contribute to the dynamics of the
fluids in a nontrivial and important way, but these issues are left
for future studies. We note however that including a radiation gas and
Thomson scattering would bring the model to a form more accurately
describing the plasma at pertinent epochs, as well as bringing it
closer to Harrison's protogalaxy model.

\ack

We like to thank Christos Tsagas for helpful discussions and
comments. This research was supported by NRF (South Africa) and Sida
(Sweden). 

\appendix

\section{Some commutator expressions}

In this section, we give some useful expressions for commuting
spatial derivatives, up to first order, in the case of dust
spacetimes (see also \cite{Roy}). 
We will assume in the following that $\3nab N$ and
$X_a$ are first order quantities.
\ba
    \fl \bra{\3nab^2N}^{\cdot} = \3nab^2\dot{N}-\ffr{2}{3}\Theta
    \3nab^2N\ , \\
    \fl \3nab_{[a}\3nab_{b]}\3nab_cN =
    \bra{\ffr{1}{9}\Theta^2-\ffr{1}{3}\mu}\3nab_{[a}N h_{b]c}\ ,\\
    \fl \3nab_{a}\3nab^2N = \3nab^2\3nab_aN +
    \bra{\ffr{2}{9}\Theta^2-\ffr{2}{3}\mu}\3nab_{a}N\ ,\\
    \fl \3nab_{[a}\3nab_{b]}\3nab^cX^d =
    \bra{-\ffr{1}{9}\Theta^2+\ffr{1}{3}\mu}
    \bras{h^c_{\ [a}\3nab_{b]}X^d + h^d_{\ [a}\3nab^cX_{b]}} \ ,\\
    \fl \3nab_a\3nab^2X_b = \3nab^2\3nab_aX_b
    +\bra{\ffr{2}{9}\Theta^2-\ffr{2}{3}\mu} \bras{\3nab_aX_b
    +\3nab_bX_a -h_{ab}\3nab_cX^c} \ ,\\
    \fl S\varepsilon^{abc} \3nab_b\3nab^2X_c =
    \3nab^2(S\varepsilon^{abc}\3nab_bX_c)\ .
\ea

\section{Isolating scalar and vector modes} \label{split}

We define our
harmonics as eigenfunctions of the Laplace--Beltrami operator
\cite{Dunsby-Bruni-Ellis,bi:harrison}
 \be \3nab^{2}Q = -\ffr{k^2}{S^2}Q,\,\,\, \dot{Q}= 0 , \label{Beltrami} \ee
where Q stands for a scalar, vector or tensor harmonic. For
example, a first order vector field $X^a$ may be expanded
covariantly in scalar and vector harmonics
\be X^a=X_S Q^a_S + X_V Q^a_V, \ee
where an implicit summation in this expansion is understood. In order to
extract the purely scalar modes, $X_S$, of a first order vector field, one
basically takes the divergence (multiplied by the scale factor $S$ for
convenience) and readily obtains the following relations:
\ba
     X \equiv S \3nab_bX^b = X_S\,(kQ_S),\label{X}
    \\ \dot{X}= S \3nab_b\dot{X}^b = \dot{X}_S\,(kQ_S),
    \\ \ddot{X} = S \3nab_b\ddot{X}^b = \ddot{X}_S\,(kQ_S).
\ea

The solenoidal modes, $X_V$, can be obtained by applying $\c$ and
noting that the curly harmonics $\tilde{Q}^a_V \equiv S
\epsilon^{abc}\3nab_bQ^V_c$ also satisfy relation \reff{Beltrami}.
The relations analogous to the scalar case are now
\ba
    \tilde{X}^a \equiv S \,\c \,X^a \;\;=
    X_V\tilde{Q}^a_V,\label{Xv}
    \\ \dot{\tilde{X}}^{\la a \ra} = S\, \c \,\dot{\tilde{X}}^{\la a \ra}
    =
    \dot{X}_V\tilde{Q}^a_V,
    \\ \ddot{\tilde{X}}^{\la a \ra} = S\, \c \,\ddot{\tilde{X}}^{\la a \ra}
    =    \ddot{X}_V\tilde{Q}^a_V.
\ea Thus, for fixed comoving wave number k, $X$ and $X_S$ as well
as $\tilde{X}^a$ and $X_V$ will obey identical equations. We like
to stress that all relations above are valid within the limits of
our two-parameter approximation scheme.

\section*{References}


\end{document}